\documentclass[fleqn,twoside]{article}
\usepackage{espcrc2}

\usepackage{graphicx}
\usepackage[figuresright]{rotating}

\usepackage{rotate}
\usepackage{epsfig}
\def\bge{\begin{equation}}
\def\ene{\end{equation}}
\newcommand{\nc}{\newcommand}
\nc{\non}{\nonumber}
\def\be{\begin{equation}}
\def\ee{\end{equation}}
\def\bga{\begin{eqnarray}}
\def\ena{\end{eqnarray}}
\def\eea{\end{eqnarray}}
\def\bg{\begin{eqnarray}}
\def\en{\end{eqnarray}}

\def\ra{\rightarrow}

\def\hbar{\not\!h}

\newcommand{\AmS}{{\protect\the\textfont2
  A\kern-.1667em\lower.5ex\hbox{M}\kern-.125emS}}

\title{Chiral Extrapolation of Hadronic Observables}

\author{A.~W.~Thomas\addressmark\thanks{University of Adelaide
preprint: ADP-02-82/T521; \hspace{1cm}
Invited presentation at The XX Int. Symposium on Lattice Field Theory,
Lattice 2002, MIT Boston, June 24--29, 2002}
\address[CSSM]{Special Research Centre for the
Subatomic Structure of Matter \\and Department of Physics and
Mathematical Physics \\
University of Adelaide, Adelaide SA 5005, Australia}}
       
\begin{document}

\begin{abstract}
One of the great challenges of lattice QCD is to produce
unambiguous predictions for the properties of physical hadrons. We
review recent progress with respect to a major barrier to achieving this
goal, namely the fact that computation time currently limits us to large
quark mass. Using insights from the study of the lattice data itself,   
together with the general constraints  
of chiral symmetry, we demonstrate that it is
possible to extrapolate accurately and in an essentially model
independent manner from the mass region where calculations
will be performed within the next five years to the chiral limit.
\vspace{1pc}
\end{abstract}

\maketitle

\section{INTRODUCTION}

Involving as it does a finite grid of
space-time points, lattice QCD \cite{Rothe:kp,Thomas:2001kw} 
requires numerous extrapolations before
one can compare with any measured hadron property.  The continuum
limit, $a \rightarrow 0$ (with $a$ the lattice spacing), 
is typically under good control \cite{Richards:2001bx}. 
With improved quark and gluon actions the ${\cal O} (a)$ 
errors can be   
eliminated so that the finite--$a$ errors are quite small, even at a modest
lattice spacing -- say 0.1 fm \cite{Zanotti:2002ax,Alford:1996pk}.  

In contrast, the infinite volume limit is
much more difficult to implement as the volume, and hence the
calculation time, scales like $N^4$. 
This limit is also inextricably
linked to the third extrapolation, namely the continuation to
small quark masses (the ``chiral extrapolation'').  The reason is, of
course, that chiral symmetry is spontaneously broken in QCD, with the
pion being a massless Goldstone boson in the chiral limit.  As the
lattice volume must contain the pion cloud of whatever hadron is under
study, one expects that the box size, $L$, should be at least  
$4 m_\pi^{-1}$….  At the
physical pion mass this is a box 5.6 fm on a side,
or a $56^4$ lattice with $a=0.1$ fm.
This is roughly $2^4$ times as big as the lattices currently in
use.

Because the time for calculations with dynamical fermions (i.e. including
quark-anti-quark creation and annihilation in the vacuum) scales 
as $m_q^{-3.6}$ \cite{Lippert:zq},  
current calculations have been limited to light quark
masses 6--10 times larger than the physical ones.
With the next generation of
supercomputers, around 10 Teraflops, it should be possible
to get as low as 2--3 times the physical quark mass, but to actually reach
that goal on an acceptable volume will require at least 500
Teraflops.  This is 10-20 years away.

Since a major motivation for lattice QCD must be to unambiguously
compare the calculations of hadron properties with experiment, this is
somewhat disappointing.  The only remedy for the next decade at least is
to find a way to extrapolate masses, form-factors, and so on, calculated
at a range of masses considerably larger than the physical ones, to the
chiral limit.  In an effort to avoid theoretical bias this has usually
been done through low-order polynomial fits as a function of quark
mass. Unfortunately, as we discuss in sect. 2, this is incorrect 
and can yield quite misleading
results because of the Goldstone nature of the pion.

Once chiral symmetry is spontaneously broken, as we have known for
decades that it must be in QCD and as confirmed in lattice
calculations, all hadron properties receive contributions involving
Goldstone boson loops.  These loops inevitably lead to results that
depend on either logarithms or odd powers of the pion mass. The
Gell-Mann-Oakes-Renner relation, however, implies that $m_\pi$ is proportional
to the square root of $m_q$, so logarithms and odd powers of $m_\pi$ are
{\it non-analytic} in the quark mass \cite{Li:1971vr}, 
with a branch point at $m_q=0$.  
One simply
cannot make a power series expansion about a branch point.

On totally general grounds, one is therefore compelled to incorporate
the non-analyticity into any extrapolation procedure.  The classical
approach to this problem is chiral perturbation theory, an effective
field built upon the symmetries of QCD \cite{Gasser:1983yg}.  
There is considerable evidence
that the scale naturally associated with chiral symmetry breaking in
QCD, $\Lambda_{\chi {\rm SB}}$, is of order $4 \pi f_\pi$, or about 1 GeV.  
Chiral perturbation theory ($\chi$pt) then
leads to an expansion in powers of $m_\pi/\Lambda_{\chi {\rm SB}}$ and  
$p/\Lambda_{\chi {\rm SB}}$, with $p$ a typical momentum
scale for the process under consideration.  At ${\cal O}(p^4)$,
the corresponding
effective Lagrangian has only a small number of unknown coefficients
which can be determined from experiment.  On the other hand, 
at ${\cal O}(p^6)$ there
are more than 100 unknown parameters 
\cite{Fearing:1994ga}, far too many to determine
phenomenologically.
\subsection{Convergence of $\chi$pt ?}
Another complication, not often discussed, is that there is yet
another mass scale entering the study of nucleon (and other baryon)
structure \cite{Detmold:2001hq,Thomas:1982kv,Donoghue:1998bs}.  
This scale is the inverse of the size of the nucleon, 
$\Lambda \sim R^{-1}$. 
Since $\Lambda$ is naturally more like a few hundred MeV,
rather than a GeV, the natural
expansion parameter, $m_\pi/\Lambda$, is of order unity for 
$m_\pi \sim 2-3 m_\pi^{\rm phys}$ -- the lowest mass scale at
which lattice data exists.  This is much larger than  
$m_\pi/\Lambda_{\chi {\rm SB}} \sim 0.3-0.4$, which
might have given one some hope for convergence.  As it is, the large
values of $m_\pi/\Lambda$ at which lattice data exist 
make any chance of a reliable
expansion in traditional (dimensionally-regulated) 
chiral perturbation theory ($\chi$pt) fairly 
minimal \cite{Hatsuda:tt}. 
Even though one has reason to doubt the
practical utility of $\chi$pt, the lattice data itself does give us some
valuable hints as to how the dilemma might be resolved.  The key is to
realize that, even though the masses may be large, one is actually
studying the properties of QCD, not a model.  In particular, one can use
the behaviour of hadron properties as a function of mass to obtain
valuable new insights into hadron structure.

\subsection{Where the Constituent Quark Picture Might Work}

The first thing that stands out, once one views the data as a whole, is
just how smoothly every hadron property behaves in the region of large
quark mass.  In fact, baryon masses behave like $a + b m_q$, magnetic moments
like $(c + d m_q)^{-1}$, charge radii squared like $(e + f m_q)^{-1}$
and so on. Thus, if one defined
a light ``constituent quark mass'' as $M \equiv M_0 + \tilde{c} m_q$
(with $\tilde{c} \sim 1$), one would find baryon masses
proportional to $M$ (times the number of u and d quarks), magnetic moments
proportional to $M^{-1}$  and so on - just as in the constituent quark picture. 
There is little or no evidence for the rapid, non-linearity
associated with the branch cuts created by Goldstone boson loops. 
Indeed, there is little evidence for a statistically significant
difference between properties calculated in quenched versus full QCD! 
How can this be?

The natural answer is readily found in the additional scale, 
$\Lambda \sim R^{-1}$, mentioned
earlier.  In QCD (and quenched QCD), 
Goldstone bosons are emitted and absorbed by
large, composite objects built of quarks and gluons.  Whenever a
composite object emits or absorbs a probe with finite momentum one must
have a form-factor which will suppress such processes for momenta
greater than $\Lambda \sim R^{-1}$. Indeed, for $m_\pi > \Lambda$
we expect Goldstone boson loops to be
suppressed as powers of $\Lambda/m_\pi$, not $m_\pi/\Lambda$ (or $m_\pi/
\Lambda_{\chi \rm SB}$).  Of course, this does not necessarily 
mean that one cannot in principle carry through the program of $\chi$pt. 
However, it does mean that there may be considerable correlations
between higher order coefficients and that it may be much more efficient
to adopt an approach which exploits the physical insight we just
explained.

Over the past three years or so we have 
developed an efficient technique to
extrapolate every hadron property which can be calculated on the lattice
from the large mass region to the physical quark mass -- while
preserving the most important non-analytic behaviour of each of those
observables.  This task is not trivial, in that various observables need
different phenomenological treatments.  On the other hand, there is a
unifying theme.  That is, pion loops are rapidly suppressed for pion
masses larger than $\Lambda$ ($m_\pi > 0.4-0.5$ GeV).  
In this region the constituent
quark model seems to represent the lattice data extremely well. 
However, for $m_\pi$ below 0.4--0.5 GeV the Goldstone loops lead to rapid,
non-analytic variation with $m_q$ and it is crucial to preserve the correct
leading non-analytic (LNA) and sometimes the next-to-leading
non-analytic (NLNA) behaviour of $\chi$pt.

In order to guide the construction of an effective, phenomenological
extrapolation formula for each hadron property, we have found it
extremely valuable to study the behaviour in a  
particular chiral quark model
-- the cloudy bag model (CBM) 
\cite{Thomas:1982kv,Miller:1979kg,Theberge:1981mq}.  
Built in the early 80's it combined a
simple model for quark confinement (the MIT bag) with a perturbative
treatment of the pion cloud necessary to ensure chiral symmetry.  The
consistency of the perturbative treatment was, not surprisingly in view
of our earlier discussion, a consequence of the suppression of high
momenta by the finite size of the pion source (in this case the bag). 
Certainly the bag model, with its sharp, static surface, has its
quantitative defects.  Yet the model can be solved in closed form and
all hadron properties studied carefully over the full range of masses
needed in lattice QCD.  Provided one works to the appropriate order the
CBM preserves the exact LNA and NLNA behaviour of QCD in the low mass
region while naturally suppressing the Goldstone boson loops for 
$m_\pi > \Lambda \sim R^{-1}$. Finally, it  
actually yields quite a good description of lattice data in
the large mass region.

We now summarise the particular situation with respect to the chiral
extrapolation of lattice data for some phenomenologically significant
baryon properties.

\section{HADRON MASSES}
By far the most extensive and accurate data for hadron properties
concerns their masses. At present the dynamical quark 
data is limited to quark masses a
little over 5 times larger than the physical light quark masses, with
most of the data at masses at least 10 times larger.
One can expect to have data at 2--3 times the physical quark mass within 5
years and hence the crucial issues for chiral extrapolation are:
\begin{itemize}
\item a) How accurately can one extract physical hadron masses from
current data?
\item b) How accurately can one extract physical hadron masses from
the next generation of data.
\item c) How model dependent are the results?
\end{itemize}

Until the developments in Adelaide in the last few 
years\cite{Leinweber:1999ig}, the usual
approach to chiral extrapolation was to draw a straight line through the
data. That is, taking the nucleon as the classic example:
\be
m_N(m_\pi) = c_0 + c_2 m_\pi^2.
\label{eq:stline}
\ee
{}For pion masses above 1 GeV or so this does not give such a good fit.
However, this can be attributed to the fact that $m_\pi^2$ is no longer
proportional to $m_q$ in this 
region \cite{Hatsuda:tt,Maris:1997tm,Maris:2000zf}. We are not concerned
with such large quark masses, instead we concentrate on the region
$m_\pi < 1$ GeV. Only recently has data in this region exhibited
significant deviation from Eq. (\ref{eq:stline}) 
\cite{Aoki:1999ff,Bernard:2001av} and this led to the more sophisticated
extrapolation function:
\be
m_N(m_\pi) = c_0 + c_2 m_\pi^2 + c_3 m_\pi^3 ,
\label{eq:cubic}
\ee
where the presence of the $m_\pi^3$ term is motivated by chiral symmetry.
(In terms of $m_q$ this term is non-analytic, being proportional
$m_q^{3/2}$, and is the LNA term in the formal expansion of the nucleon
mass.) Indeed, it is relatively simple to see that $c_3$ comes from the
pion pole term in the $\pi$N self-energy process and hence that it is
{\it model independent} \cite{Li:1971vr}:
\be
c_3 \equiv c_{\rm LNA} = - \frac{3 g_A^2}{32 \pi f_\pi^2} , 
\label{eq:c3}
\ee
with $g_A$ and $f_\pi$ the axial charge of the nucleon and the pion
decay constant, respectively, in the chiral limit.

The first empirical indication of serious problems in this approach came
with the realization that a fit to lattice data gives $c_3 \sim -0.76$
GeV$^{-2}$, whereas the {\it model independent} value  given by 
$\chi$pt, Eq.(\ref{eq:c3}), is -5.6 GeV$^{-2}$ -- a factor of 8 larger!
This tells us immediately that {\bf either} there are serious convergence
problems with Eq. (\ref{eq:cubic}) {\bf or} lattice QCD is in error.
Clearly most readers would opt for the first possibility and so do we.

The formal series for $m_N$ about the chiral SU(2) limit is usually
written as 
\bga 
&&m_N = m_0 + c_2 m_\pi^2 + c_{\rm LNA} m_\pi^3 + c_4 m_\pi^4 
\non \\
&& +c_{\rm NLNA} m_\pi^4 \ln m_\pi + c_5 m_\pi^5 + c_6 m_\pi^6 +.....
\label{eq:formal}
\ena
where the next-to-leading non-analytic term, $m_\pi^4 \ln m_\pi$, is
dominated by the N $\rightarrow \Delta \pi \rightarrow$ N loop. However,
{}for any meaningful extrapolation of lattice data in the next decade
this expression is essentially useless. It is derived in the limit
$m_\pi << \Delta (\equiv m_\Delta - m_N)$, whereas the lowest lattice
data with dynamical fermions that one can expect in the next decade is
perhaps 200-250 MeV -- c.f. $\Delta = 292$ MeV. In fact, most lattice
data will still lie above $\Delta$. Mathematically the region around
$m_\pi \approx \Delta$ is dominated by a square root 
branch cut which starts at $m_\pi = \Delta$. Using dimensional
regularization this takes the form \cite{Banerjee:1995wz}:
\bga
&& \frac{6 g_A^2}{25 \pi^2 f_\pi^2}\left[ (\Delta^2 - m_\pi^2)^{\frac{3}{2}}
\ln (\Delta + m_\pi -\sqrt{\Delta^2 - m_\pi^2}) \right. \non \\
&& \left. - \frac{\Delta}{2}
(2 \Delta^2 - 3 m_\pi^2) \ln m_\pi \right],
\label{eq:log}
\ena
for $m_\pi < \Delta$, while for $m_\pi > \Delta$ the first logarithm
becomes an arctangent. No serious attempt has been made to extend the
formal expansion in Eq. (\ref{eq:formal}) to incorporate this cut in an
analysis of lattice data and,  
given the number of parameters to be determined if one works to order
$m_\pi^6$, it is not likely that it will be done in the next decade.

Even ignoring the $\Delta \pi$ cut for a short time, studies of the
formal expansion of the N~$\ra $N~$\pi \ra $~N
self-energy integral ($\sigma_{{\rm N} \pi}$, suggest that it has
abysmal convergence properties. Using a sharp, ultra-violet cut-off,
Wright showed \cite{SVW} that the series diverged for $m_\pi > 0.4$ GeV.
If one instead uses a dipole cut-off, which in view of the
phenomenological shape of the nucleon's axial form-factor is much more
realistic, it is worse -- with the radius of convergence being around
0.25 GeV.

In summary, the lattice data itself, formal studies in the NJL model
and phenomenological studies of the nucleon based on long-distance
regularization all suggest that the radius of convergence of the formal
chiral expansion of the nucleon mass barely touches the lowest data
point that we will get from lattice QCD in the next decade. This is not
a productive approach to the problem.

\subsection{The Solution}
On the other hand, there is a better way. One can directly use 
the  form \cite{Leinweber:1999ig}:
\be
m_N = \alpha + \beta m_\pi^2 + \sigma_{{\rm N} \pi}(m_\pi, \Lambda) + 
\sigma_{\Delta \pi}(m_\pi, \Lambda),
\label{eq:extrap}
\ee
where $\sigma_{{\rm B} \pi}$ is the self-energy arising from a B $\pi$
loop (B = N or $\Delta$) 
and $\alpha, \beta$ and $\Lambda$ are determined by fitting {\it
lattice data}. For the reasons outlined, it is essential that
the self-energies are evaluated using some
ultra-violet regulator -- a sharp cut-off or a dipole form, for
example. Whatever is chosen does not effect the non-analytic structure
which is guaranteed correct. The branch points at $m_\pi$ equals zero and 
$\Delta$ are incorporated naturally. While $\alpha$ and $\beta$ {\bf
will} depend on the regulator, one can easily expand the self-energy
terms to order $m_\pi^2$ (or higher) to obtain the chiral coefficients
at the appropriate order to compare with effective field theory -- e.g.
see Ref. \cite{Donoghue:1998bs} for a full discussion of this issue.
The essential point is that studies of the nucleon, the $\Delta$ (c.f.
Fig. 4 of Leinweber {\it et al.} \cite{Leinweber:1999ig}) and the $\rho$
meson \cite{Leinweber:2001ac} suggest that {\it this procedure will  
result in little or no model dependence in the
extrapolation to the physical pion mass once there is accurate lattice data
for $m_\pi \sim 0.3$ GeV or less}. Physically this is possible because
the self-energy loops are rapidly suppressed in the region $m_\pi > 0.4$
GeV. Thus, an extrapolation based on Eq. (\ref{eq:extrap}) allows one
to respect  {\it all the chiral constraints}, keep the number of fitting
parameters low and yield essentially model independent results at the
physical pion mass. No other approach can do this.

\subsection{A Possible Connection to QQCD}
The study of baryon spectroscopy in quenched lattice QCD (QQCD) has
recently made great progress. We have already noted that the lattice
data behaves like a constituent quark model for quark masses above
50--60 MeV because Goldstone boson loops are strongly suppressed in this
region. This not only provides a very natural explanation of the
similarity of quenched and full data in this region but it also suggests
a much more ambitious approach to hadron spectra. It suggests that one
might remove the small effects of Goldstone boson loops in QQCD
(including the $\eta'$) and then estimate the hadron masses in full QCD
by introducing the Goldstone loops which yield the LNA and NLNA
behaviour in full QCD.

As a first test of this idea, Young {\it et al.} \cite{Young:2002cj} 
recently analysed the
MILC data \cite{Bernard:2001av} for the N and $\Delta$, using  
Eq.(\ref{eq:extrap}) for full QCD and 
the appropriate generalization for QQCD -- i.e. using quenched
pion couplings as well as the single and double $\eta'$ 
loops \cite{Labrenz:1996jy,Young:rx}. The
results were remarkable, with the values of $\alpha$ and $\beta$ for the
N (or the $\Delta$) obtained in QQCD agreeing within statistical errors
with those obtained in full QCD. Certainly this result is somewhat
dependent on the shape of the ultra-violet cut-off chosen -- although
the extent of that is yet to be studied in detail. Nevertheless, given
that the study involved the phenomenologically favoured dipole
form, it is a remarkable result and merits further investigation.

\section{ELECTROMAGNETIC PROPERTIES OF HADRONS}
\begin{figure}[htb]
\begin{center}
\includegraphics[angle=90,height=13pc]{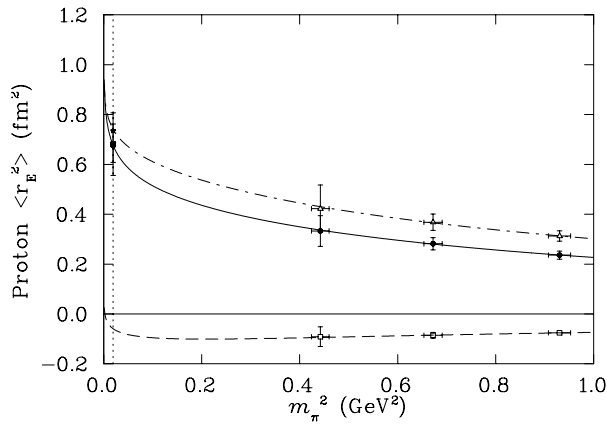}
\caption{Chiral extrapolation of
lattice results for the squared electric charge radius
of the proton (solid line) 
-- from Ref. \protect\cite{Hackett-Jones:2000js}. Fits
to the contributions from individual quark flavors are also
shown (the $u$-quark results are indicated by open triangles
and the $d$-quark results by open squares). The physical value
predicted by the fit is indicated at the physical pion mass, while 
the experimental value is denoted by an asterisk.}
\label{fig:prot}
\end{center}
\end{figure}

Although there is only limited lattice data for
hadron charge radii, recent experimental progress in the determination
of hyperon charge radii \cite{Adamovich:pe},
has led us to examine the extrapolation
procedure for extracting charge radii from the lattice 
simulations.  Figure \ref{fig:prot}  
shows the extrapolation of the
lattice data for the charge radius of the proton, including the $\ln
m_\pi$ (LNA) term in a generalised Pad\'e 
approximant~\cite{Hackett-Jones:2000js,Dunne:2001ip}:
\be 
<r^2>_{\rm ch}^p = \frac{c_1 + \chi \ln \frac{m_\pi^2}{m_\pi^2 +
\mu^2}}{1 + c_2 m_\pi^2}.
\ee
Here $c_1$ and $c_2$ are parameters determined by fitting the lattice
data in the large mass region ($m_\pi^2 > 0.4$ GeV$^2$), while $\mu$,
the scale at which the effects of pion loops are suppressed, is not yet
determined by the data but is simply set to 0.5 GeV. The coefficient
$\chi$ is model independent and determined by chiral perturbation
theory. Clearly the 
agreement with experiment is much better if, as shown, 
the logarithm required by
chiral symmetry is correctly included -- rather than simply
making a linear extrapolation in the quark mass (or $m_\pi^2$).
{}Full details of the results for all the octet baryons may be found in
Ref. \cite{Hackett-Jones:2000js}.

The situation for baryon magnetic moments is also very interesting.
The LNA contribution in this case arises from the diagram where
the photon couples to the pion loop.  As this
involves two pion propagators the expansion of the proton and neutron
moments is:
\be
\mu^{p(n)} = \mu^{p(n)}_0 \mp \alpha m_\pi + {\cal O}(m_\pi^2).
\label{eq:10}
\ee
Here $\mu^{p(n)}_0$ is the value in the chiral limit and the
linear term in $m_\pi$ is proportional to $m_q^{\frac{1}{2}}$,
a branch point at $m_q = 0$.  The
coefficient of the LNA term is $\alpha = 4.4 \mu_N $GeV$^{-1}$.
At the physical pion mass this LNA
contribution is $0.6\mu_N$, which is almost a third of the neutron
magnetic moment \cite{Leinweber:2001ui}.

\begin{figure}[htb]
\begin{center}
\includegraphics[angle=90,height=13pc]{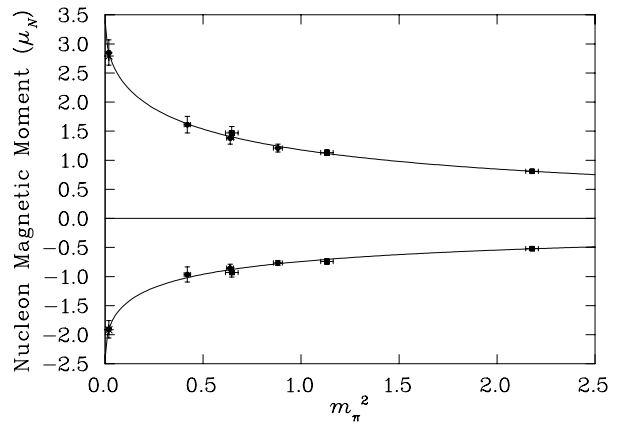}
\caption{Chiral extrapolations of lattice QCD magnetic moments 
{}for the proton (upper) and
neutron (lower).  The curves illustrate a
two parameter fit, Eq. (\protect\ref{eq:11}),
to the simulation data, using a Pad\'e approximant, in which
the one-loop corrected chiral coefficient of $m_\pi$ is taken from
$\chi$pt.  The experimentally measured moments are indicated by
asterisks. The figure is taken from  
Ref.~\protect\cite{Leinweber:1999ej}.}
\label{NucleonMomFit}
\end{center}
\end{figure}

As for the charge radii, 
the chiral behaviour of $\mu^{p(n)}$ is vital for a correct
extrapolation of lattice 
data. From  Fig.~\ref{NucleonMomFit} we see that
one can obtain a very satisfactory fit to
some rather old data (which happens to be the best available)
using the simple Pad\'e approximant \cite{Leinweber:1999ej}:
\be
\mu^{p(n)} = \frac{\mu^{p(n)}_0}{1 \pm \frac{\alpha}{\mu^{p(n)}_0} m_\pi
+
\beta m_\pi^2} .  
\label{eq:11}
\ee
Existing lattice data can only determine two parameters and Eq.(\ref{eq:11})
has just two free parameters while guaranteeing the correct LNA
behaviour (i.e. the correct value of $\alpha$) 
as $m_\pi \ra 0$ {\it as well as} the correct behaviour of HQET
at large $m_\pi^2$.  The
extrapolated values of $\mu^p$ and $\mu^n$
at the physical pion mass, $2.85 \pm 0.22 \mu_N$ and $-1.90 \pm 0.15
\mu_N$ are currently the best estimates from non-perturbative QCD
\cite{Leinweber:1999ej}. (Similar results, including NLNA terms in
chiral perturbation theory, have been reported recently by Hemmert and Weise 
\cite{Hemmert:2002uh}.) For the application
of similar ideas to other members of the
nucleon octet we refer to Ref. \cite{Hackett-Jones:2000qk}, while for the
strangeness magnetic moment of the
nucleon we refer to Ref. \cite{Leinweber:2000nf}.
The last example is another case where there have been 
tremendous improvements in the experimental capabilities.  
Specifically, the accurate measurement of
parity violation in $ep$ scattering 
\cite{Kumar:2000eq}
is giving us vital information on
hadron structure.
\begin{figure}[htb]
\begin{center}
\includegraphics[height=20pc]{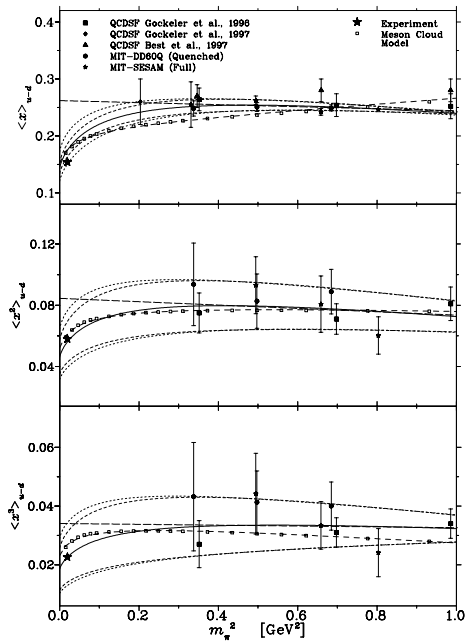}
\caption{Moments of the $u - d$ quark distribution from various lattice
simulations.
The straight (long-dashed) lines are linear fits
to this data, while the curves have the correct LNA behaviour in
the chiral limit -- see the text for details.
The small squares are the results of the meson cloud model
and the dashed curve through them
best fits using Eq.~(\protect\ref{eq:fit}).  The stars represent the
phenomenological values taken from NLO fits in
the $\overline{\rm MS}$ scheme. The figure is taken from
Ref.~\protect\cite{Detmold:2001jb}.}
\label{mainfig}
\end{center}
\end{figure}

In concluding this section, we note that the observation that chiral
corrections are totally suppressed for $m_\pi$ above about 0.4 GeV and
that the lattice data looks very like a constituent quark picture there
suggests a novel approach to modelling hadron structure. It seems that
one might avoid many of the complications of the chiral quark models, as
well as many of the obvious failures of constituent quark models by
building a new constituent quark model with $u$ and $d$ masses in the
region of the strange quark -- where SU(3) symmetry should be exact.
Comparison with data could then be made after the same sort of chiral
extrapolation procedure that has been applied to the lattice data.
Initial results obtained by Cloet {\it et al.} for the octet baryon magnetic
moments using this approach are very promising indeed \cite{Cloet:2002eg}.
We note also the extension to $\Delta$-baryons, including the NLNA
behaviour, reported for the first time in 
Ref.~\cite{Cloet2}.

\section{MOMENTS OF STRUCTURE FUNCTIONS}
The moments of the parton distributions measured in lepton-nucleon
deep inelastic scattering   
are related, through the operator product expansion,
to the forward nucleon matrix elements of certain local twist-2 operators
which can be accessed in lattice simulations~\cite{Thomas:2001kw}.
The more recent data, used in the present analysis, are taken from the
QCDSF \cite{Gockeler:1997jk} and MIT 
\cite{Dolgov:2001ca} groups and shown in Fig.~\ref{mainfig} for the
$n=1$, 2 and 3 moments of the $u - d$ difference
at NLO in the $\overline{\rm MS}$ scheme.

To compare the lattice results with the experimentally measured
moments, one must extrapolate in quark mass from about  
50 MeV to the physical value.
Naively this is done by assuming that the moments depend linearly on the
quark mass.  However, as shown in Fig.~\ref{mainfig} (long dashed
lines), a linear extrapolation of the world lattice data for the $u-d$
moments typically overestimates the experimental values by 50\%.  This
suggests that important physics is still being omitted from the
lattice calculations and their extrapolations.

Here, as for all other hadron properties, a linear extrapolation
in $\bar m \sim m_\pi^2$ must fail as it omits crucial nonanalytic
structure associated with chiral symmetry breaking.  
The leading nonanalytic (LNA) term for the $u$ and
$d$ distributions in the physical nucleon arises from the single pion
loop dressing of the bare nucleon and has been
shown \cite{Thomas:2000ny,LNA_DIS} to
behave as $m_\pi^2 \log m_\pi$.
Experience with the chiral behaviour of masses and magnetic moments
shows that the LNA terms alone are not sufficient to describe lattice
data for $m_\pi > 0.2$~GeV.  Thus, in order to fit the 
lattice data at larger $m_\pi$, while preserving the correct chiral
behaviour of moments as $m_\pi \to 0$, a low order, analytic expansion
in $m_\pi^2$ is also included in the extrapolation and the moments of
$u-d$ are fitted with the form \cite{Detmold:2001jb}:
\bga
\langle x^n \rangle_{u-d}
&=& \ a_n\ +\ b_n\ m_\pi^2\
\non \\ &+& \ a_n\ c_{\rm LNA}\ m_\pi^2
\ln \left( \frac{m_\pi^2}{m_\pi^2 + \mu^2} \right)\ ,
\label{eq:fit}
\ena
where the coefficient,
$c_{\rm LNA} = -(3 g_A^2+1)/(4\pi f_\pi)^2$ \cite{LNA_DIS}. 
The parameters $a_n$,  
$b_n$ are determined by fitting the lattice data.  The mass $\mu$
determines the scale above which pion loops no longer yield rapid
variation and corresponds to the upper limit of the momentum
integration if one applies a sharp cut-off in the pion loop integral.
Consistent with our earlier discussion of this scale it
is taken to be 0.55 GeV.
Multi-meson loops and other contributions cannot give rise to LNA
behaviour and thus, near the chiral limit,  
Eq.~(\ref{eq:fit}) is the most general form for moments of the PDFs at
${\cal O}(m_\pi^2)$ which is consistent with chiral symmetry.

We stress that $\mu$ is not yet determined by the
lattice data and it is indeed possible
to consistently fit both the lattice
data and the experimental values with $\mu$ ranging from
0.4~GeV to 0.7~GeV. This dependence on $\mu$ is illustrated in
Fig.~\ref{mainfig} by the difference between the inner and outer
envelopes on the fits.  Data at   
smaller quark masses, ideally $m_\pi^2 \sim$ 0.05--0.10 GeV$^2$,
are therefore crucial to constrain this parameter
in order to perform an accurate extrapolation based solely on lattice data.

Although we do not have sufficient space to explore the matter in
detail, we note that once we have a parametrization of the first four
moments as a function of pion mass it is possible to study the behaviour
of the PDFs as a function of $x$  and $m_\pi$ \cite{Detmold:2001dv}. 
Remarkably, for $m_\pi >
0.5$ GeV the distribution tends to peak at $x \sim 1/3$,
resembling very much what one would expect in a constituent quark model.
This further reinforces the discussion in sect.~1.2 concerning the
behaviour of hadron structure within QCD as the light quark mass exceeds
50--60 MeV.
\subsection{Spin--Dependent Moments}
There has been considerable interest in spin-dependent structure
functions since the discovery of the spin crisis by EMC. As a result we
now have a great deal of experimental information on spin-dependent 
PDFs \cite{DATAREVIEW}. More recently
studies of the axial charge of the nucleon within lattice QCD have
revealed a surprisingly strong dependence on the volume of the 
lattice \cite{Sasaki:2001th}. Of course, interest is not limited to the
first moment of the usual PDFs. One can calculate higher moments on
the lattice, as well as moments of the transverse structure functions
and experiments are also planned to explore the latter.

The analysis of the chiral extrapolation of moments of the unpolarized
PDFs in the previous subsection has recently been extended to the
polarized sector by Detmold {\it et al.} \cite{Detmold:2002nf}. 
Years of phenomenological experience suggest that the $\Delta$ resonance
might be important for spin-dependent observables. Within the CBM the
explicit appearance of the $\Delta$ played a vital role in the
convergence of a perturbative expansion of nucleon properties, including
its axial charge \cite{Dodd:1981ve}. Thus it is not surprising that the
explicit inclusion of the $\Delta$ in the extrapolation of the
spin-dependent moments has a very significant effect. Whereas the rapid
variation of the unpolarized moments is unaffected, the
inclusion of the $\Delta$ almost totally removes the non-linearity for the
polarized moments. While Detmold {\it et al.} suggest a new  
extrapolation formula which accurately approximates the non-analytic
behaviour in this case, one could fairly reliably use a linear chiral
extrapolation for the spin-dependent moments. For a full discussion of
the results we refer to Ref. \cite{Detmold:2002nf}, simply noting here
that with a careful extrapolation of data taken on a relatively large
lattice \cite{SchierholzPC}
the agreement between theory and experiment for $g_A$ is now
quite reasonable: $g_A({\rm Latt}) = 1.12 \pm 0.05$ compared with
$g_A({\rm Expt.}) = 1.267$. 

\section{CONCLUSION}

At the present time we have a wonderful conjunction of opportunities.
Modern accelerator facilities are
providing data of unprecedented precision over a tremendous kinematic
range at the same time as numerical simulations of lattice QCD are
delivering results of impressive accuracy.  It is therefore timely to
ask how to use these advances to develop a new and deeper understanding
of hadron structure and dynamics.

In combination with the very successful techniques for
chiral extrapolation, which we have illustrated by just a few examples,
lattice QCD will finally yield accurate data on the consequences of
non-perturbative QCD. Furthermore, the physical insights obtained from
the study of hadron properties as a function of quark mass will
guide the development of new quark models and hence a much more
realistic picture of hadron structure.

\section*{ACKNOWLEDGEMENTS}
I would like to thank those colleagues who have
contributed to my understanding of the problems discussed here,
notably Ian Cloet, Will Detmold, Derek Leinweber,
Wally Melnitchouk, Stewart Wright and Ross Young.
I would also like to thank Steven Bass and Ross Young for their comments
and suggestions on this manuscript.
This work was supported by the Australian Research Council and the
University of Adelaide.


\begin{thebibliography}{0}
%
\bibitem{Rothe:kp}
H.~J.~Rothe,
World Sci.\ Lect.\ Notes Phys.\  {\bf 59} (1997) 1.
%
\bibitem{Thomas:2001kw}
A.~W.~Thomas and W.~Weise,
``The Structure of the Nucleon,''
{\it  289 pages. Hardcover ISBN 3-527-40297-7 Wiley-VCH, Berlin 2001}.
%
\bibitem{Richards:2001bx}
D.~G.~Richards {\it et al.}, 
Nucl.\ Phys.\ Proc.\ Suppl.\  {\bf 109}, 89 (2002).
%
\bibitem{Zanotti:2002ax}
J.~M.~Zanotti {\it et al.},
Nucl.\ Phys.\ Proc.\ Suppl.\  {\bf 109}, 101 (2002).
%
\bibitem{Alford:1996pk}
M.~G.~Alford, T.~R.~Klassen and G.~P.~Lepage,
Nucl.\ Phys.\ Proc.\ Suppl.\  {\bf 53}, 861 (1997).
%
\bibitem{Lippert:zq}
T.~Lippert, S.~Gusken and K.~Schilling,
Nucl.\ Phys.\ Proc.\ Suppl.\  {\bf 83}, 182 (2000).
%
\bibitem{Li:1971vr}
L.~F.~Li and H.~Pagels,
Phys.\ Rev.\ Lett.\  {\bf 26}, 1204 (1971).
%
\bibitem{SVW}
S.~V.~Wright, Ph. D. thesis (The University of Adelaide, 2002).
%
\bibitem{Gasser:1983yg}
J.~Gasser and H.~Leutwyler,
Annals Phys.\  {\bf 158}, 142 (1984).
%
\bibitem{Fearing:1994ga}
H.~W.~Fearing and S.~Scherer,
Phys.\ Rev.\ D {\bf 53}, 315 (1996)
[arXiv:hep-ph/9408346].
%
\bibitem{Detmold:2001hq}
W.~Detmold {\it et al.},
Pramana {\bf 57}, 251 (2001)
[arXiv:nucl-th/0104043].
%
\bibitem{Thomas:1982kv}
A.~W.~Thomas,
Adv.\ Nucl.\ Phys.\  {\bf 13}, 1 (1984).
%
\bibitem{Donoghue:1998bs}
J.~F.~Donoghue, B.~R.~Holstein and B.~Borasoy,
Phys.\ Rev.\ D {\bf 59}, 036002 (1999).
%
\bibitem{Hatsuda:tt}
T.~Hatsuda,
Phys.\ Rev.\ Lett.\  {\bf 65}, 543 (1990).
%
\bibitem{Miller:1979kg}
G.~A.~Miller, A.~W.~Thomas and S.~Theberge,
Phys.\ Lett.\ B {\bf 91}, 192 (1980).
%
\bibitem{Theberge:1981mq}
S.~Theberge, G.~A.~Miller and A.~W.~Thomas,
Can.\ J.\ Phys.\  {\bf 60}, 59 (1982).
%
\bibitem{Leinweber:1999ig}
D.~B.~Leinweber, A.~W.~Thomas, K.~Tsushima and S.~V.~Wright,
Phys.\ Rev.\ D {\bf 61}, 074502 (2000)
[arXiv:hep-lat/9906027].
%
\bibitem{Banerjee:1995wz}
M.~K.~Banerjee and J.~Milana,
Phys.\ Rev.\ D {\bf 54}, 5804 (1996).
%
\bibitem{Maris:1997tm}
P.~Maris and C.~D.~Roberts,
Phys.\ Rev.\ C {\bf 56}, 3369 (1997)
[arXiv:nucl-th/9708029].
%
\bibitem{Maris:2000zf}
P.~Maris,
arXiv:nucl-th/0009064.
%
\bibitem{Aoki:1999ff}
S.~Aoki {\it et al.}  [CP-PACS Collaboration],
Phys.\ Rev.\ D {\bf 60}, 114508 (1999).
%
\bibitem{Bernard:2001av}
C.~W.~Bernard {\it et al.},
Phys.\ Rev.\ D {\bf 64}, 054506 (2001)
[arXiv:hep-lat/0104002].
%
\bibitem{Leinweber:2001ac}
D.~B.~Leinweber {\it et al.},
Phys.\ Rev.\ D {\bf 64}, 094502 (2001)
[arXiv:hep-lat/0104013].
%
\bibitem{Young:2002cj}
R.~D.~Young {\it et al.},
arXiv:hep-lat/0205017.    
%
\bibitem{Labrenz:1996jy}
J.~N.~Labrenz and S.~R.~Sharpe,
Phys.\ Rev.\ D {\bf 54}, 4595 (1996)
[arXiv:hep-lat/9605034].
%
\bibitem{Young:rx}
R.~D.~Young {\it et al.},
Nucl.\ Phys.\ Proc.\ Suppl.\  {\bf 109A} (2002) 55.
%
\bibitem{Adamovich:pe}
M.~I.~Adamovich {\it et al.}  [WA89 Collaboration],
Eur.\ Phys.\ J.\ C {\bf 8} (1999) 59.
%
\bibitem{Hackett-Jones:2000js}
E.~J.~Hackett-Jones {\it et al.},
Phys.\ Lett.\ B {\bf 494}, 89 (2000)
[hep-lat/0008018].
%
\bibitem{Dunne:2001ip}
G.~V.~Dunne {\it et al.},
Phys. Lett. {\bf B531}, 77 (2002)
[hep-th/0110155].
%
\bibitem{Leinweber:2001ui}
D.~B.~Leinweber {\it et al.},
Phys.\ Rev.\ Lett.\  {\bf 86}, 5011 (2001).
[arXiv:hep-ph/0101211].
%
\bibitem{Leinweber:1999ej}
D.~B.~Leinweber {\it et al.},
Phys.\ Rev.\ D {\bf 60}, 034014 (1999).
%
\bibitem{Hemmert:2002uh}
T.~R.~Hemmert and W.~Weise,
arXiv:hep-lat/0204005.
%
\bibitem{Hackett-Jones:2000qk}
E.~J.~Hackett-Jones {\it et al.},
Phys.\ Lett.\ B {\bf 489}, 143 (2000).
%
\bibitem{Leinweber:2000nf}
D.~B.~Leinweber and A.~W.~Thomas,
Phys.\ Rev.\ D {\bf 62}, 074505 (2000).
%
\bibitem{Kumar:2000eq}
K.~S.~Kumar and P.~A.~Souder,
Prog.\ Part.\ Nucl.\ Phys.\  {\bf 45}, S333 (2000).
%
\bibitem{Cloet:2002eg}
I.~C.~Cloet {\it et al.}, 
Phys.\ Rev.\ C {\bf 65}, 062201 (2002)
[arXiv:hep-ph/0203023].
%
\bibitem{Cloet2}
I.~C.~Cloet {\it et al.},
in Proc. of the Joint Workshop on ``Physics at
Japanese Hadron Facility'', World Scientific (2002), to appear.
%
\bibitem{Gockeler:1997jk}
M.~Gockeler {\it et al.},
Nucl.\ Phys.\ Proc.\ Suppl.\  {\bf 53}, 81 (1997).
%
\bibitem{Dolgov:2001ca}
D.~Dolgov {\it et al.},
Nucl.\ Phys.\ Proc.\ Suppl.\  {\bf 94}, 303 (2001).
%
\bibitem{Thomas:2000ny}
A.~W.~Thomas {\it et al.}, 
Phys.\ Rev.\ Lett.\  {\bf 85}, 2892 (2000)
[arXiv:hep-ph/0005043].
%
\bibitem{LNA_DIS}
D.~Arndt and M.~J.~Savage,
Nucl.\ Phys.\ A {\bf 697}, 429 (2002); 
J.~W.~Chen and X.~d.~Ji,
Phys.\ Lett.\ B {\bf 523}, 107 (2001).
%
\bibitem{Detmold:2001jb}
W.~Detmold {\it et al.},
Phys.\ Rev.\ Lett.\  {\bf 87}, 172001 (2001)
[arXiv:hep-lat/0103006].
%
\bibitem{Detmold:2001dv}
W.~Detmold {\it et al.},
Eur.\ Phys.\ J.\ directC {\bf 13}, 1 (2001)
[arXiv:hep-lat/0108002].
%
\bibitem{DATAREVIEW}
M.~Erdmann,
Talk given at 8th International Workshop on Deep Inelastic Scattering
and QCD (DIS 2000), Liverpool, England, 25-30 Apr. 2000,
arXiv:hep-ex/0009009;
%
\bibitem{Sasaki:2001th}
S.~Sasaki {\it et al.},
Nucl.\ Phys.\ Proc.\ Suppl.\  {\bf 106}, 302 (2002).
[arXiv:hep-lat/0110053].
%
\bibitem{Detmold:2002nf}
W.~Detmold, W.~Melnitchouk and A.~W.~Thomas,
arXiv:hep-lat/0206001.
%
\bibitem{SchierholzPC}
G.~Schierholz,
at 9th Int. Conf. on Baryon Structure,  
Jefferson Lab, March 2002;
%
M.~Gockeler {\it et al.},
arXiv:hep-ph/9909253;
%
and \emph{private communication}.
%
\bibitem{Dodd:1981ve}
L.~R.~Dodd {\it et al.},
Phys.\ Rev.\ D {\bf 24}, 1961 (1981).
%
\end{thebibliography}
\end{document}